\documentclass[12pt,a4paper]{article}

\usepackage{cite}
\usepackage{latexsym,amssymb,amsmath,amscd,amsthm,amsxtra,graphicx}

\newcommand{\ba}{\begin{eqnarray}}
\newcommand{\ea}{\end{eqnarray}}
\newcommand{\bc}{\begin{center}}
\newcommand{\ec}{\end{center}}

\newcommand{\ovl}{\overline} 
\newcommand{\chiP}{\chi_{{}_P}^{}}
\newcommand{\barchiP}{\ovl\chi_{{}_P}^{}}
\newcommand{\tildechiP}{\widetilde\chi_{{}_P}^{}}

\newcommand{\lvac}{\langle 0 |}
\newcommand{\rvac}{| 0 \rangle}

\begin{document}

\title{Studying the scalar bound states of the $K\ovl K$ system in the
Bethe-Salpeter formalism}
\author{\\Xin-Heng Guo\thanks{xhguo@bnu.edu.cn} \, and
Xing-Hua Wu\thanks{wuxh@brc.bnu.edu.cn} \\
{\small\it Institute of Low Energy Nuclear Physics, Beijing Normal University},
\\
{\small\it Beijing 100875, China}
}
\date{}
\maketitle

\thispagestyle{empty}
\begin{abstract}

We study the possible bound states of the $K\ovl K$ system in the
Bethe-Salpeter formalism in the ladder and instantaneous
approximations. We find that the bound states exist. However,
these bound states have very small decay widths. Therefore,
besides the possible $K\ovl K$ component, there
may be some other structures in the observed $f_0(980)$ and $a_0(980)$\,.\\

\end{abstract}

\newpage
\setcounter{page}{1}

\section{Introduction}

Although the dynamics of quarks and gluons at the low energy scale is expected
to be relativistic and strongly coupled, the simple non-relativistic quark
model sucessfully describes the properties of most light mesons
($q\bar q$) and baryons ($qqq$). However, exception occurs for
some scalar particles. Just as stated in Ref. \cite{gn-rmp-99}, ``the features
of QCD {\it not} ({\it all}) contained in the (simple) quark model''.
To describe these overpopulated scalar particles, non-$q\bar q$
structures have been assigned to them for about three decades since the study
in Ref. \cite{jaffe-prd-77}. They have been regarded as four-quark states
\cite{jaffe-prd-77,ai-npb-89,as-prd-98,a-ph-0309118}\,, or molecules
composed of conventional particles
\cite{wi-prd-83-90,ldhs-npa-90,jphs-prd-95,oller-npa-2003,kls-prd-2004,zcsz-prd-2006} (e.g. $K\ovl K$ for $f_0(980)$ and/or $a_0(980)$), etc..
Up to now, the puzzle about the nature of these scalar particles still remains
unsolved. For example, to describe
the recent experimental data \cite{a-ex-98-00} and the more accurate
measurement by KLOE Collaboration \cite{kloe-2002}, $f_0(980)$ and/or
$a_0(980)$ were regarded as four-quark states \cite{a-ph-0309118} or molecular
binding of $K\ovl K$ \cite{oller-npa-2003} and both of them lead to results
consistent with the experiments. Obviously, further investigation
on the structure of these scalar particles is necessary.


On the other hand, more and more overpopulated states (especially those
containing heavy flavors) have been discovered and confirmed by various
experiments \cite{belle-cdfII-d0-babar}. Due to the proximity
of these particles' masses to those of two lowest lying conventional particles
(carrying certain heavy flavor(s)), one would naturally identify them as
molecules of conventional particles (see, e.g. Refs.
\cite{heavy-molecules,zcsz-prd-2006}). Therefore, it is interesting to study
whether this picture about these scalar systems is right or not.

In this paper we will focus on the scalar particles $f_0(980)$ and $a_0(980)$.
One purpose of the present paper is to investigate whether the bound
states of the $K\ovl K$ system, interacting by exchanging various vector
particles ($\rho$, $\omega$, $\phi$), exist. The other purpose is to discuss
the extent to which the $K\ovl K$ component contributes to the observed
particles $f_0(980)$ and $a_0(980)$\,.

We choose the Bethe-Salpeter (BS) formalism (in the ladder
approximation and the instantaneous approximation) as our starting
point. The main reason for this is that, in comparison with the
potential model (used in e.g.
Refs.\cite{kls-prd-2004,zcsz-prd-2006}), one can include some
relativistic corrections automatically in the BS equation.

One may wonder whether the ladder approximation taken for our
pseudo-scalar system in this paper is suitable. In fact, there
have been some works in which the legitimacy of the application of
the ladder approximation in the BS formalism has been studied, see
e.g. Refs.
\cite{gross-prc-26-2203,itzykson-zuber-books,nieuwenhuis-tjon-prl-77-814,theussl-desplanques-fbs-30-5,barrobergflodt-etal-fbs-2006,emamirazavi-etal-jpg-2006}.
For example, in Ref. \cite{gross-prc-26-2203} it was shown that
including only ladder graphs in the scalar-scalar system can not
lead to the correct one-body limit. Furthermore, in the gauge
theory, within the ladder approximation gauge invariance can not
be maintained. To solve these problems, at least crossed-ladder
graphs should be included
\cite{gross-prc-26-2203,itzykson-zuber-books}. More recently, it
was shown that the crossed-ladder graphs do contribute large
corrections to the ladder approximation in some cases
\cite{nieuwenhuis-tjon-prl-77-814,theussl-desplanques-fbs-30-5}.
For large enough coupling, the contribution from the
crossed-ladder graphs becomes even more important than that from
the ladder ones.

In our case, the square of the effective dimensionless coupling
constant (see Eqs. (\ref{lintkkrho})-(\ref{lintkkphi}) and
(\ref{kernel-t-channel}) in Sect. \ref{sect-bound-state-of-KK})
can be written as $g_{KK{\rm V}}^2E^2 / (4\pi M_{\rm K}^2)$ which
is greater than 3
     \footnote{
       In our case, the binding energy is
       of order {$\cal O$}($10^1$) MeV, the total energy of the binding
       system is $E\approx 2 M_{\rm K}$\,, and the coupling
       $g_{KK{\rm V}}$ is about 3 when ${\rm V}=\rho,\omega$ and about
       $-3\sqrt{2}$ when ${\rm V}=\phi$.
     }.
From the n\"aive point of view, for such a large coupling
constant, the ladder approximation is not legitimate
\cite{theussl-desplanques-fbs-30-5}. However, a closer examination
shows that there is a significant difference between our case and
the cases discussed in Refs.
\cite{nieuwenhuis-tjon-prl-77-814,theussl-desplanques-fbs-30-5}
(see also Refs.
\cite{barrobergflodt-etal-fbs-2006,emamirazavi-etal-jpg-2006}), in
which the mass of the exchanged particle is very small. On the
contrary, the exchanged particles, $\rho$, $\omega$, $\phi$, in
our case have large masses compared with the constituent particles
$K$ and $\ovl K$\,. We will show that the large masses of the
exchanged particles suppress significantly the contribution of the
crossed-ladder graphs since factors of the form $1/(p^2-M_{\rm
V}^2)$ from the extra propagators in the crossed-ladder graphs
lead to extra suppression (in powers of $1/M_{\rm V}^2$).
Therefore, in our case, the net contribution of the crossed-ladder
graphs is in fact very small compared with that of the ladder
graphs (more details are given in Sect.
\ref{sect-bound-state-of-KK}).

Since the contribution of the crossed-ladder graphs is small in our case
the problems associated with the ladder approximation, if existing, will
not be serious.

Another approximation we will take is the instantaneous
approximation. In this approximation, the energy exchanged between
the constituent particles of the binding system is neglected. This
is appropriate if the relativistic effects in the system are
small. Our calculations (in the ladder approximation and the
instantaneous approximation) show that both the iso-scalar and
iso-vector bound states of the $K\ovl K$ system with small binding
energy exist. This shows that the binding of the constituent
particles is weak, hence the exchange of energy between them can
be neglected.

However, regarding these bound states as the observed particles,
$f_0(980)$ and $a_0(980)$, one can find that the decay widths of these bound
states are too small to explain the experimental data. In other words, while
the bound states of $K\ovl K$ do contribute to the observed scalar particles,
they themselves can not describe the full properties of these particles.

The remainder of this paper is organized as follows. In Sect. \ref{sect-BS},
we review the BS formalism for the system of two pseudo-scalar particles and
discuss the normalization condition of the BS wave function. In
Sect. \ref{sect-bound-state-of-KK}, we discuss the bound state equations for
the $K\ovl K$ system in detail. The decays of the $K\ovl K$ bound state
to $\pi\pi$ and $\pi\eta$ final states are discussed in
Sect. \ref{sect-decay-width-of-KK}\,. The numerical results are presented in
Sect. \ref{sect-numerical}\,. The final section is reserved for some
discussions and our conclusions.

\section{The Bethe-Salpeter formalism}\label{sect-BS}

In this section we will review the general formalism of the BS equation and
derive the BS equation for the system of two pseudo-scalar particles.
We will also derive the normalization condition for the BS wave function.
Let us start by defining the BS wave function for the bound state $|P\rangle$
of two pseudo-scalar particles as the following:
\ba
\label{definition-BS}
\chiP(x_1,x_2) = \langle 0 | {\rm T}\,\phi_1(x_1)\phi_2(x_2) | P \rangle
= e^{-iPX} \chiP(x)\,,
\ea
where $\phi_1(x_1)$ and $\phi_2(x_2)$ are the field operators of two
pseudo-scalar particles, respectively, $P$ denotes the total momentum of the
bound state, and the relative coordinate $x$ and the center of mass coordinate
$X$ are defined by
\ba
X=\eta_1 x_1 + \eta_2 x_2\,,\quad x = x_1 - x_2 \,,
\ea
or inversely,
\ba
x_1 = X + \eta_2 x\,,\quad x_2 = X - \eta_1 x \,,
\ea
where $\eta_i = m_i/(m_1+m_2)$\,, $m_i\, (i=1,2)$ is the mass of the $i$-th
constituent particle. The equation for the BS wave function can be derived
from a four-point Green function,
\ba
S(x_1,x_2;y_2,y_1)
= \lvac {\rm T}\,\phi_1(x_1)\phi_2(x_2) (\phi_1(y_1)\phi_2(y_2))^\dag \rvac \,.
\label{four-point-green-function}
\ea
To obtain the BS equation, we express the above four-point Green
function in terms of the four-point truncated irreducible kernel $\ovl K$\,,
\ba
&&S(x_1,x_2;y_2,y_1) = S_{(0)}(x_1,x_2;y_2,y_1) \nonumber\\
&& \quad + \int d^4u_1 d^4 u_2 d^4 v_1 d^4 v_2\, S_{(0)}(x_1,x_2;u_2,u_1)
{\ovl K} (u_1,u_2;v_2,v_1) S(v_1,v_2;y_2,y_1) \,,
\label{irreducible-rep}
\ea
where $S_{(0)}$ is related to the forward scattering disconnected four-point
amplitude,
\ba
S_{(0)}(x_1,x_2;y_2,y_1) = \Delta_1(x_1,y_1)\Delta_2(x_2,y_2) \,,
\ea
where $\Delta_i(x_i, y_i)$ is the complete propagator of the $i$-th particle,
\ba
\Delta_i(x, y) = \lvac {\rm T}\,\phi_i(x)\phi_i(y)^\dag \rvac
=\int {d^4k\over(2\pi)^4} \, e^{-ik(x-y)}\,\Delta_i(k,m_i) \,.
\ea
From Eqs. (\ref{definition-BS}) and (\ref{irreducible-rep}) one can derive the
following BS equation for the bound state of two pseudo-scalar particles:
\ba
\chiP(x_1,x_2)
= \int d^4 u_1 d^4 u_2 d^4 v_1 d^4 v_2\, S_{(0)}(x_1,x_2;u_2,u_1)
{\ovl K} (u_1,u_2;v_2,v_1) \, \chiP(v_1,v_2) \,,
\ea
or, by inverting $S_{(0)}$,
\ba
\int d^4 y_1 d^4 y_2 \, S_{(0)}^{-1}(x_1,x_2;y_2,y_1) \chiP(y_1,y_2)
= \int d^4 v_1 d^4 v_2\, {\ovl K} (x_1,x_2;v_2,v_1) \, \chiP(v_1,v_2) \,.
\label{bs-equation}
\ea
In this paper, we will investigate the BS equation in momentum space, in which
the BS wave function is obtained as (using Eq. (\ref{definition-BS}))
\ba
\chiP(p_1,p_2) = \int d^4 x_1 d^4 x_2\, e^{ip_1x_1 + ip_2x_2}\chiP(x_1,x_2)
=(2\pi)^4\,\delta(p_1+p_2-P) \chiP(p) \,,
\ea
where $p = \eta_2p_1 - \eta_1p_2$ is the relative momentum and
$\chiP(p) = \int d^4x \, e^{ipx}\chiP(x)$\,.
The Fourier transformation of the four-point Green function in
Eq. (\ref{four-point-green-function}) reads
\ba
S(x_1,x_2;y_2,y_1) = \int {d^4P d^4P' d^4p d^4p'\over (2\pi)^{16}}
e^{-iPX+iP'Y-ipx+ip'y} \, \widetilde S(p,p',P,P') \,,
\ea
with $\widetilde S(p,p',P,P') = (2\pi)^4\delta^4(P-P')\widetilde S_P(p,p')$.
Similarly, for the irreducible kernel we have
\ba
\ovl K(x_1,x_2;y_2,y_1) = \int {d^4P d^4P' d^4p d^4p'\over (2\pi)^{16}}
e^{-iPX+iP'Y-ipx+ip'y} \, \ovl K(p,p',P,P') \,,
\ea
with $\ovl K(p,p',P,P') = (2\pi)^4\delta^4(P-P')\ovl K_P(p,p')$.
The relative momenta and the total momentum of the bound state in the
equations are defined by
\ba
p = \eta_2p_1 - \eta_1p_2\,,\quad p' = \eta_2p'_1 - \eta_1p'_2\,,\quad
P = p_1 + p_2 = p'_1 + p'_2 \,,
\label{momentum-transform}
\ea
or inversely,
\ba
p_1 = \eta_1 P + p\,,\quad p_2 = \eta_2 P - p\,, \quad
 p_1' = \eta_1 P + p'\,,\quad p_2' = \eta_2 P - p' \,.
\label{momentum-transform-inverse}
\ea
We must note that the constituents of the bound state can not be on-shell,
otherwise the bound state is not a really bound state. Consequently,
$p_i^2\neq m_i^2$ (and similar for $p'_i$).

Then, the inhomogeneous equation (\ref{irreducible-rep}) in momentum space
reads
\ba
\label{inhomogeneous-momentum}
\int {d^4k \over (2\pi)^4}\left[I_P(p,k)+\ovl K_P(p,k)\right]
\widetilde S_P(k,p') = (2\pi)^4\delta(p-p')\,,
\ea
where $I_P(p,k) = -(2\pi)^4 \delta^4(p-k)
\Delta_1^{-1}(p_1,m_1)\Delta_2^{-1}(p_2,m_2)$\,.
The BS equation (\ref{bs-equation}) for the bound state in momentum space
takes the following form:
\ba
\label{homogeneous-momentum}
\int {d^4k \over (2\pi)^4}\left[I_P(p,k)+\ovl K_P(p,k)\right] \chiP(k) = 0 \,.
\ea
This is a homogeneous equation for the BS wave function.

From the BS bound state equation, Eq. (\ref{bs-equation}) in coordinate
space or Eq. (\ref{homogeneous-momentum}) in momentum space, we can
see that the BS wave function satifies a homogeneous equation. Therefore,
its normalization can not be determined from the bound state equation.
To obtain the correct normalization of the BS wave function,
following Ref.\cite{lurie-book}, we start by considering the contribution
of the bound state with $P^0=E_{\bf P}$ to the four-point function
$S$. Let us first isolate the contributions from some possible bound states.
Consider the case with $\min\{x_1^0,x_2^0\}>\max\{y_1^0,y_2^0\}$ and
insert a complete set of states into the four-point Green function, we have
\ba
S(x_1,x_2;y_2,y_1)
&\makebox[0pt]{=}& \sum_{\bf P}
\lvac {\rm T}\,\phi_1(x_1)\phi_2(x_2) |{ P}\rangle
\langle { P}|{\rm T}\, \phi_2(y_2)^\dag\phi_1(y_1)^\dag \rvac
\Big|_{\min\{x_1^0,x_2^0\}>\max\{y_1^0,y_2^0\}}
\nonumber\\
&\makebox[0pt]{=}& \int {d^3{\bf P}\over (2\pi)^3} \,
 e^{-iE_{\bf P}(X^0-Y^0)+i{\bf P}\cdot({\bf X}-{\bf Y})} \chiP(x)\barchiP(y)
\Big|_{\min\{x_1^0,x_2^0\}>\max\{y_1^0,y_2^0\}} \,.
\label{bound-state-rep}
\ea
Furthermore, the requirement ${\min\{x_1^0,x_2^0\}>\max\{y_1^0,y_2^0\}}$
can be described by a theta-function,
\ba
\theta\left(X^0-Y^0+{\eta_2-\eta_1\over 2}(x^0-y^0)-{|x^0|\over 2}
-{|y^0|\over 2}\right) \,.
\ea
Using this representation and the contour-integral definition of the
theta-function, Eq. (\ref{bound-state-rep}) can be written as
\ba
S(x_1,x_2;y_2,y_1) &=& {i} \int {d^4 P\over (2\pi)^4} \,
 e^{i{\bf P}\cdot({\bf X}-{\bf Y})-iP^0(X^0-Y^0)} \chiP(x)\barchiP(y)
\, {1\over P^0-E_{\bf P}+i\epsilon}
\nonumber\\&&\quad\times\,
e^{-i(P^0-E_{\bf P})
\left[(\eta_2-\eta_1)(x^0-y^0)-|x^0|-|y^0|\right]/2} \,.
\ea
Hence, near the pole at $P^0=E_{\bf P}$, we have
\ba
\label{pole-representation}
\widetilde S_P(p,p') = {i \over P^0-E_{\bf P}+i\epsilon}\,
\chiP(p)\barchiP(p')
+ \hbox{terms regular at } P^0=E_{\bf P} \,.
\ea
Now, define an auxiliary quantity:
\ba
Q_P(p,p')=\int {d^4k\over (2\pi)^4}\, (P^0-E_{\bf P})\widetilde S_P(p,k)
{\partial\over\partial P^0}\left[I_P(k,p')+\ovl K_P(k,p')\right] \,.
\ea
For convenience, we can imagine the arguments of $\ovl K_P$, $I_P$,
$\widetilde S_P$\,, and $Q_P$ as matrix indices and write the above quantity
in a compact form:
\ba
Q_P = (P^0-E_{\bf P})\widetilde S_P {\partial\over\partial P^0}
\left[I_P+\ovl K_P\right] \,.
\ea
In terms of this notation we can also rewrite Eqs.
(\ref{inhomogeneous-momentum}), (\ref{homogeneous-momentum}),
and (\ref{pole-representation}) as
\ba
 \widetilde S_P \left[I_P+\ovl K_P\right] &=& 1 \,, \\
 \left[I_P+\ovl K_P\right]\chiP &=& 0\,,\qquad (P^0=E_{\bf P}) \,, \\
 \lim_{P^0\to E_{\bf P}}(P^0-E_{\bf P})\widetilde S_P
&=& i\,\chiP\barchiP \,.
\ea
Using the above equations and operating $Q_P$ upon $\chiP$ we have the
normalization condition for the BS wave function,
\ba
\label{normalization-condition}
i\int {d^4p\, d^4p'\over (2\pi)^8} \,\barchiP(p){\partial\over\partial P^0}
\left[I_P(p,p')+\ovl K_P(p,p')\right]\chiP(p')=1 \,,\quad
P^0 = E_{\bf P} \,.
\ea

The BS equation (\ref{homogeneous-momentum}) is very complex. Without
approximation we can not even write down the irreducible kernel and the
propagators of particles explicitly. Since the binding of the $K\ovl K$
system is weak we use the so-called instantaneous approximation:
${\ovl K}_P(p,p')={\ovl K}_P({\bf p},{\bf p}')$. Furthermore, the propagator
is set to have the form of the free one. Then, the BS equation
(\ref{homogeneous-momentum}) becomes
\ba
-(p_1^2-m_1^2)(p_2^2-m_2^2)\chiP(p)
= \int {d^4p'\over (2\pi)^4} {\ovl K}_P({\bf p},{\bf p}')\chiP(p')\,.
\label{bs-equation-momentum}
\ea
Now, we divide Eq. (\ref{bs-equation-momentum}) by the two propagators on
both sides and then perform the integration over $p^0$ and $p'{}^0$.
Then we have
\ba
\label{3-dim-BS}
{E^2-(E_1+E_2)^2\over (E_1+E_2)/E_1E_2} \tildechiP({\bf p})
={i\over 2}\int{d^3p'\over(2\pi)^3}\,
{\ovl K}_P({\bf p},{\bf p}')\tildechiP({\bf p}') \,,
\ea
where $E_i \equiv \sqrt{{\bf p}^2 + m_i^2}$, $E=P^0$\,, and the equal-time
wave function is defined as
\ba
\tildechiP({\bf p})= \int dp^0 \, \chiP(p) \,.
\ea
While deriving Eq. (\ref{3-dim-BS}) we have used the following result (in the
rest frame of the bound state):
\ba
\int_{-\infty}^\infty\,
{dp^0\over (p_1^2 - m_1^2 + i\epsilon)(p_2^2 - m_2^2+ i\epsilon)}
= -i\pi\, { (E_1+E_2)/E_1E_2\over E^2-(E_1+E_2)^2}  \,,
\ea
which can be obtained by choosing a proper contour. For convenience we define
the following potential:
\ba
\label{kernel-potential}
V({\bf p},{\bf p}') = {i\over E_1 E_2 (E_1+E_2)}\, {\ovl K}_P({\bf p},{\bf p}')
\,.
\ea
Then, the BS bound state equation can be written as
\ba
\label{scalar-BS-eq-final}
\left[{E^2\over (E_1+E_2)^2} -1 \right] \tildechiP({\bf p})
= {1\over 2}\int{d^3p'\over(2\pi)^3}\, V({\bf p},{\bf p}')\tildechiP({\bf p}')
\,.
\ea
Eq. (\ref{scalar-BS-eq-final}) will be the starting point in our later
numerical calculations.

For later convenience we also write out $\chiP(p)$ in terms of
$\tildechiP({\bf p})$\,. From Eqs. (\ref{bs-equation-momentum}) and
(\ref{3-dim-BS}) we have
\ba
\chiP(p^0,{\bf p})
&=& {-1\over (p_1^2-m_1^2+i\epsilon)(p_2^2-m_2^2+i\epsilon)}
\int{d^3p'\over(2\pi)^4}\, {\ovl K}_P({\bf p},{\bf p}')\tildechiP({\bf p}')
\nonumber\\
&=& {i\over \pi}\,{1\over (p_1^2-m_1^2+i\epsilon)(p_2^2-m_2^2+i\epsilon)}\,
{E^2-(E_1+E_2)^2\over (E_1+E_2)/E_1E_2} \tildechiP({\bf p}) \,,
\label{4-dim-BS}
\ea
where $p_1^2-m_1^2=(\eta_1 E+p^0)^2-E_1^2$\,,
$p_2^2-m_2^2=(\eta_2 E-p^0)^2-E_2^2$\,.

\section{The bound state(s) of the $K\ovl K$ system}
\label{sect-bound-state-of-KK}

In this section, we will study the possible bound state of the $K\ovl K$
system. The lowest lying particles with the strangeness numbers $\pm 1$ form two
isospin doublets: $(K^+ , K^0)^{\rm T}$ and $(-\ovl K^{\, 0}, K^-)^{\rm T}$\,,
where the superscript T denotes transpose
       \footnote{
      The conventions about the isospin multiplets used here
      and in the following are the same as those used in e.g.
      Ref. \cite{neville-pr-67}\,.
    }.
One can gather them into two fields, $K_1$ and $K_2$\,, which have the
following expansion in momentum space
\ba
\label{field-expand-1}
K_1 &=& \int {d^3 p\over (2\pi)^3}{1\over \sqrt{2 E_{\bf p}}}
(a_{K^+}^{}e^{-ipx}+a_{K^-}^\dag e^{ipx}) \,,
\\
\label{field-expand-2}
K_2 &=& \int {d^3 p\over (2\pi)^3}{1\over \sqrt{2 E_{\bf p}}}
(a_{K^0}^{}e^{-ipx}+a_{\ovl K^{\,0}}^\dag e^{ipx}) \,,
\ea
where $E_{\bf p}=\sqrt{{\bf p}^2+m_K^2}$ is the energy of the particles and
we omit the effects of isospin violation so that the masses of the kaons are
the same. These two fields can be grouped into an isospin doublet
$K=(K_1, K_2)^{\rm T}$\,, which furnishes the fundamental representation of
the isospin group $SU(2)_f$\,.

The $K\ovl K$ system has isospin 1 or 0. The iso-scalar bound state can be
written as
\ba
|P\rangle_{0}
= {1\over\sqrt{2}}\left| K^+K^- + K^0\ovl K{}^{\,0} \right\rangle \,,
\ea
and the three components of the iso-vector states are
\ba
|P\rangle_{1,0}
= {1\over\sqrt{2}} \left| K^+ K^- - K^0\ovl K{}^{\,0} \right\rangle\,,\quad
|P\rangle_{1,+1} = - \left| K^+ \ovl K{}^{\,0}\right\rangle\,,\quad
|P\rangle_{1,-1} = \left| K^- K^0\right\rangle \,.
\ea
Let us now project the bound states on the field operators $K_1$ and $K_2$.
From Eqs. (\ref{field-expand-1}) and (\ref{field-expand-2}) we have
\ba
\label{projection-of-BS-state}
\langle 0| {\rm T}\,\{ K_i(x_1) K_j(x_2)^\dag \} |P\rangle_{I,I_3}
= C_{(I,I_3)}^{ij}\,\chiP {}^{(I)}(x_1,x_2) \,,
\ea
where $\chiP {}^{(I)}$ is the common BS wave function for the bound state with
isospin $I$ which depends only on the state $|P\rangle_{I,I_3}$ (as will be
shown later, the BS wave function only depends on $I$ but not on $I_3$) but
not on the concrete field contents. The isospin coefficients $C^{ij}_{(I,I_3)}$
for the iso-scalar state are
\ba
\label{isoscalar-coefficient-0}
C_{(0,0)}^{11}=C_{(0,0)}^{22} = 1/\sqrt{2}\,,\qquad \hbox{else } = 0 \,,
\ea
and for the iso-vector state we have
\ba
\label{isovector-coefficient-1}
C_{(1,0)}^{11}=-C_{(1,0)}^{22} = 1/\sqrt{2}\,,\quad
C_{(1,+1)}^{12} = -1\,,\quad C_{(1,-1)}^{21} = 1
\,,\qquad \hbox{else } = 0 \,.
\ea
Now consider the kernel. The BS equation
(\ref{bs-equation-momentum}) for the bound state can be written down
schematically,
\ba
\Delta_1^{-1}\Delta_2^{-1}C_{(I)}^{ij}\chiP {}^{(I)}
= {\ovl K}_P^{\,ij,lk}C_{(I)}^{kl}\chiP {}^{(I)} \,,
\ea
where $\Delta_{1,2}$ are the propagators of the constituent particles.
Then, from Eq. (\ref{isoscalar-coefficient-0}), for the iso-scalar case,
we have (take $ij=11$ as an example)
\ba
\Delta_1^{-1}\Delta_2^{-1}\chiP {}^{(0)}
= ({\ovl K}_P^{\,11,11}+{\ovl K}_P^{\,11,22})\chiP {}^{(0)} \,.
\ea
Similarly, for the iso-vector case, taking the $I_3=0$ component as an
example, we have
\ba
\Delta_1^{-1}\Delta_2^{-1}\chiP {}^{(1,0)}
= ({\ovl K}_P^{\,11,11}-{\ovl K}_P^{\,11,22})\chiP {}^{(1,0)} \,.
\ea
From the above equations, we can see that if the iso-scalar bound state
exists one can not ensure the existence of the iso-vector bound state,
and vice versa.

The interactions among the kaons and vector particles, $\rho$, $\omega$,
$\phi$, at the level of hadrons are described by the
$SU(3)_{\rm V}^{}\times SU(3)_{\rm A}^{}$ chiral dynamics. The relevant
interaction vertices are (see e.g. Ref. \cite{zcsz-prd-2006})
\ba
{\cal L}_{KK\rho}&=&
i g_{_{KK\rho}}^{}K^\dag  (\vec\tau\cdot \vec\rho^{\,\mu}) \partial_\mu K
+ \hbox{c.c.} \,,
\label{lintkkrho}
\\
{\cal L}_{KK\omega}&=&
i g_{_{KK\omega}}^{} K^\dag (\partial_\mu K)\omega^\mu + \hbox{c.c.} \,,
\label{lintkkomega}
\\
{\cal L}_{KK\phi}&=&
i g_{_{KK\phi}}^{} K^\dag (\partial_\mu K)\phi^\mu + \hbox{c.c.} \,,
\label{lintkkphi}
\ea
where c.c. is the complex conjugate of the first term
and $g_{_{KK{\rm V}}}^{}$ (V can be $\rho$, $\omega$, and $\phi$) are the
coupling constants which can be related to $g_{\rho\pi\pi}$
in the $SU(3)_f$ limit,
\ba
g_{_{KK\rho}}^{}= g_{\rho\pi\pi}/2 \,,\quad
g_{_{KK\omega}}^{}= g_{\rho\pi\pi}/2 \,,\quad
g_{_{KK\phi}}^{} = - g_{\rho\pi\pi}/\sqrt 2 \,.
\label{coupling-constant}
\ea
The $\rho\pi\pi$ coupling is determined by
$g_{\rho\pi\pi}=M_\rho/(\sqrt{2}f_\pi)\approx 6$ \cite{KSRF-66}\,, where
$M_\rho$ is the mass of $\rho$ and $f_\pi$ is the decay constant of the pion.

From the above observations, at the tree level, in $t$-channel we
have the following kernel for the BS equation in the so-called
ladder approximation
       \footnote{
     When the exchanging meson is $\rho$\,, we have
     ${\ovl K}_P^{\,11,22}(\rho)=2{\ovl K}_P^{\,11,11}(\rho)$ and
     ${\ovl K}_P^{\,12,12}(\rho)=-{\ovl K}_P^{\,11,11}(\rho)$\,;
     when the exchanging meson is $\omega$, we have
     ${\ovl K}_P^{\,11,22}(\omega)=0$ and
     ${\ovl K}_P^{\,12,12}(\omega)
     ={\ovl K}_P^{\,11,11}(\omega)$\,; the case for $\phi$ is the same
     as that for $\omega$\,.
       }:
\ba
&&\ovl K(p_1,p_2;p'_2,p'_1;M_{\rm V})
= -\,i(2\pi)^4\delta^4(p'_1+p'_2-p_1-p_2)
\nonumber \\[5pt] && \qquad\qquad \times ~
c_I^{} \, g_{_{KK{\rm V}}}^2
{(p_1+p_1')\cdot(p_2+p_2')+(p_1^2-p_1'{}^2)(p_2^2-p_2'{}^2)/M_{\rm V}^2
\over (p_1-p_1')^2-M_{\rm V}^2} \,,
\label{kernel-t-channel}
\ea
where $c_I^{}$ is the isospin coefficient: $c_0^{}=3,1,1$ and
$c_1^{}=-1,1,1$ for $\rho$, $\omega$, $\phi$\,,
respectively. These results are consistent with those in Ref.
\cite{ldhs-npa-90}. In Eq. (\ref{kernel-t-channel}) we have used the
following propagator for a massive vector meson:
\ba
\Delta_{\mu\nu}(p, M_{\rm V})
={-i\over p^2-M_{\rm V}^2}(g_{\mu\nu}-p_\mu p_\nu/M_{\rm V}^2) \,.
\ea
From the above analysis, we can see that the BS wave function depends only
on the isospin $I$ but not on its component $I_3$\,. This is because we
have only considered strong interactions which preserve the isospin symmetry.
Therefore, we will omit the $I_3$ label and write $\chiP {}^{(I,I_3)}$
simply as $\chiP {}^{(I)}$ from now on.

In the following discussion and calculation in this paper, we will
take the kernel (\ref{kernel-t-channel}) as our starting point
(this is the so-called ladder approximation) and will not consider
those non-ladder (e.g. crossed) graphs. Before proceeding, let us
discuss briefly the contribution of the crossed-ladder graphs. Our
aim is to estimate the ratio of their contribution to that of the
ladder graph. To make the calculation of the (4-th order) crossed
graph tractable, the following simplification are taken: terms
carrying $p^2$, $p'{}^2$, $p\cdot p'$, etc. are omitted in the
calculation. Since we will stay in the rest frame of the binding
system, we also set $p\cdot P$ and $p'\cdot P$ to zero (the
instantaneous approximation). These approximations will be
appropriate if the half width of the BS wave function is small
enough compared with the masses $M_{\rm V}$ and $M_{\rm K}$\,.
From the numerical results (in Sect. \ref{sect-numerical}), we can
see that the half width of the BS wave function is about $0.1$
GeV, which is indeed very small compared with the masses $M_{\rm
V}$ and $M_{\rm K}$\,. The physical picture for this approximation
is that the configurations with small momenta are dominant in the
model.

Taking these approximations, we calculate explicitly the
contribution from the (4-th order) crossed graph. We take the case
where the exchanged particle is $\rho$ as an example, other two
cases with $\omega$ and $\phi$ as exchanged particles give similar
results. We work with the dimensional regularization method and
the minimal subtraction scheme while calculating the crossed
graph. The numerical results show that the ratio of the
contribution from the crossed graph to that from the ladder (2nd
order) graph has the following form: \ba 0.12 - 0.01 \ln
{\mu^2\over 1 {\rm GeV}^2}\,, \ea where $\mu$ is the
renormalization scale. Since only one (but not all) higher order
graph is calculated, the result depends on this renormalization
scale. It is natural to take $\mu$ to be around 1 GeV, which is
the scale of chiral symmetry breaking. From this result, we can
see that the ratio of the contribution from the crossed graph to
that from the ladder one is less than 15\% (in the case where
$\omega$ is the exchanged particle the result is almost the same
while in the case where $\phi$ is the exchanged particle, we have
the ratio $\approx 25$\%).

For comparison, let us reduce manually the masses of the exchanged particles to
$0.15 M_{\rm K}$, which is the case discussed in Refs.
\cite{nieuwenhuis-tjon-prl-77-814,theussl-desplanques-fbs-30-5}.
We find that this ratio will rise to $300\%$ or even more:
$3.84 + 3.38 \ln {\mu^2\over 1 {\rm GeV}^2}$\,.
This result is consistent with those shown in Refs
\cite{nieuwenhuis-tjon-prl-77-814,theussl-desplanques-fbs-30-5}.

Now, let us consider only the ladder approximation and proceed by taking the
instantaneous approximation, $p^0 = 0$ and $p'{}^0=0$ in the kernel
(\ref{kernel-t-channel}), and stay in the center-of-mass frame of the bound
state, ${\bf P}=0$\,. Then, the potential in Eq. (\ref{kernel-potential}) due
to the exchange of a vector meson V becomes (using
Eq. (\ref{momentum-transform-inverse}))
\ba
V^{(I)}({\bf p},{\bf p}';M_{\rm V}) &\makebox[0pt]{=}& c_I \,U({\bf p},{\bf p}';M_{\rm V})
\nonumber\\
&\makebox[0pt]{=}& c_I\,{-g_{_{KK{\rm V}}}^2\over E_1 E_2
(E_1+E_2)}\, {({\bf p}+{\bf p}')^2 + 4\eta_1\eta_2 E^2 + ({\bf
p}^2-{\bf p}'{}^2)^2/M_{\rm V}^2 \over ({\bf p}-{\bf p}')^2 +
M_{\rm V}^2}  \,. \label{potential-with-isospin-factor} \ea In
order to describe the phenomena in the real world, we should
include a form factor at each interacting vertex of hadrons to
include the finite-size effects of these hadrons. For the meson
($q\bar q$) case, the form factor is assumed to take the following
form \cite{ldhs-npa-90}: \ba \label{form-factor} F({\bf k}) =
{2\Lambda^2 - M_{\rm V}^2 \over 2\Lambda^2 + {\bf k}^2}\,,\quad
{\bf k} = {\bf p}-{\bf p}' \,, \ea where $\Lambda$ is a cutoff
parameter which will be adjusted to give the solution of the BS
equation. At the lowest order, the BS equation includes $F^2$ in
its kernel, i.e. $V\to V\cdot F^2$. The most important term in the
numerator of Eq. (\ref{potential-with-isospin-factor}) is
$4\eta_1\eta_2 E^2$. Other terms are small since the momenta of
the constituent particles of the binding system are small. After
transforming into the form in coordinate space, similar to the
case in Ref. \cite{zcsz-prd-2006}, one can see that the potential
is in fact a Yukawa-like potential (the sum of a Yukawa potential
and several derivatives of the Yukawa potential).

Then, for the bound state of the $K\ovl K$ system, the BS equation
(\ref{scalar-BS-eq-final}) becomes
\ba
\left[{E^2\over (E_1+E_2)^2}-1\right]\,\tildechiP{}^{(I)}(|{\bf p}|)
= {1\over 2}\int{d^3p'\over(2\pi)^3}\,
V_{\rm eff}^{(I)}({\bf p},{\bf p}')F({\bf k})^2 \tildechiP{}^{(I)}(|{\bf p}'|)
\,,
\ea
where the effective potential is (depending on isospin $I$)
\ba
V_{\rm eff}^{(0)}({\bf p},{\bf p}') &=& \,3 U({\bf p},{\bf p}'; M_\rho)
+U({\bf p},{\bf p}'; M_\omega) + U({\bf p},{\bf p}'; M_\phi) \,,
\\
V_{\rm eff}^{(1)}({\bf p},{\bf p}') &=& -\, U({\bf p},{\bf p}'; M_\rho)
+U({\bf p},{\bf p}'; M_\omega) + U({\bf p},{\bf p}'; M_\phi) \,.
\ea
If we are interested in the ground state of the BS equation,
the corresponding BS wave function is in fact rotational invariant, i.e.
$\tildechiP({\bf p})$ depends only on the norm of the three momentum,
$|{\bf p}|$\,. Therefore, after completing the azimuthal integration, the
above BS equation becomes a one-dimensional-integral equation, which reads
\ba
\tildechiP{}^{(I)}(|{\bf p}|)
= \int d|{\bf p}'|\, V_{\rm 1d}^{(I)}(|{\bf p}|,|{\bf p}'|)
\tildechiP{}^{(I)}(|{\bf p}'|) \,,
\label{scalar-BS-eq-full-final}
\ea
where $V_{\rm 1d}^{(I)}(|{\bf p}|,|{\bf p}'|)$ are one-dimensional effective
potentials:
$V_{\rm 1d}^{(0)}(|{\bf p}|,|{\bf p}'|)= 3 U_{\rm 1d}(M_\rho)
+U_{\rm 1d}(M_\omega) + U_{\rm 1d}(M_\phi)$\,, and
$V_{\rm 1d}^{(1)}(|{\bf p}|,|{\bf p}'|)= -\, U_{\rm 1d}(M_\rho)
+U_{\rm 1d}(M_\omega) + U_{\rm 1d}(M_\phi)$
with
\ba
U_{\rm 1d}(M_{\rm V}) =
- {g_{_{KK{\rm V}}}^2\over 4(2\pi)^2}\,
{E_1+E_2\over E_1E_2\big[E^2-(E_1+E_2)^2\big]}\,
{|{\bf p}'|\over |{\bf p}|} (V_1+V_2+V_3) \,,
\nonumber
\ea
where
\ba
V_1 &\makebox[0pt]{=}& -4|{\bf p}||{\bf p}'|\,(2\Lambda^2 - M_{\rm V}^2)
\,{2\Lambda^2+2(2\eta_1\eta_2 E^2 + |{\bf p}|^2 + |{\bf p}'|^2)
+(|{\bf p}|^2 - |{\bf p}'|^2)^2/M_{\rm V}^2
\over \big[2\Lambda^2 + (|{\bf p}|+|{\bf p}'|)^2\big]
   \big[2\Lambda^2 + (|{\bf p}|-|{\bf p}'|)^2\big]} \,,
\nonumber\\[5pt]
V_2 &\makebox[0pt]{=}&
\Big[M_{\rm V}^2 + 2(2\eta_1\eta_2 E^2 + |{\bf p}|^2 + |{\bf p}'|^2)
+(|{\bf p}|^2 - |{\bf p}'|^2)^2/M_{\rm V}^2 \Big]
 \ln{M_{\rm V}^2 + (|{\bf p}|+|{\bf p}'|)^2 \over
M_{\rm V}^2 + (|{\bf p}|-|{\bf p}'|)^2} \,,
\nonumber\\[5pt]
V_3 &\makebox[0pt]{=}&
-\,\Big[M_{\rm V}^2 + 2(2\eta_1\eta_2 E^2 + |{\bf p}|^2 + |{\bf p}'|^2)
+(|{\bf p}|^2 - |{\bf p}'|^2)^2/M_{\rm V}^2 \Big]
\ln{2\Lambda^2 + (|{\bf p}|+|{\bf p}'|)^2\over 2\Lambda^2
+ (|{\bf p}|-|{\bf p}'|)^2} \,.
\nonumber
\ea

\section{The decay width of the $K\ovl K$ system}
\label{sect-decay-width-of-KK}

To find out the bound states of the $K\ovl K$ system, one only needs to solve
the homogeneous BS equation. However, when we want to calculate physical
quantities such as the decay width we have to face the problem of the
normalization of the BS wave function. In the following we will discuss the
normalization of the BS wave function $\tildechiP(|{\bf p}|)$\,.

Substituting the relation between $\chiP(p)$ and $\tildechiP(|{\bf p}|)$\,,
Eq. (\ref{4-dim-BS}), and Eqs. (\ref{kernel-potential})
(\ref{potential-with-isospin-factor}) into the normalization equation
(\ref{normalization-condition}) one arrives at the following normalization
equation for $\tildechiP(|{\bf p}|)$ (after carrying out some
$p_0$-integrations with proper contours):
\ba
-{1\over \pi^2}\int {d^3{\bf p}\over (2\pi)^3}
\left[\tildechiP{}^{(I)}(|{\bf p}|)\right]^2  R
- {2\eta_1\eta_2 E \over \pi^2}\int {d^3{\bf p}d^3{\bf p}'\over (2\pi)^6} \,
\tildechiP{}^{(I)}(|{\bf p}|)\tildechiP{}^{(I)}(|{\bf p}'|) \,
F^2 \, H^{(I)}=1 \,,
\label{normalization-condition-temp-3}
\ea
where $F(={2\Lambda^2 - M_{\rm V}^2 \over 2\Lambda^2 + ({\bf p}-{\bf p}')^2})$
is the form factor, $H^{(0)}=3H(M_\rho)+H(M_\omega)+H(M_\phi)$ and
$H^{(1)}=-H(M_\rho)+H(M_\omega)+H(M_\phi)$ with
$H(M_{\rm V})={g_{_{KK{\rm V}}}^2\over ({\bf p}-{\bf p}')^2+M_{\rm V}^2}$\,,
and
\ba
R&=&
-\,E \Big[ -2 E^2 (E_1^2-E_2^2) (E_1 \eta_1-E_2 \eta_2)
+E^4 (E_1 \eta_1+E_2 \eta_2)
\nonumber\\&&\qquad\qquad\quad
+\,(E_1^2-E_2^2)
(E_1^3 \eta_1+3 E_1 E_2^2 \eta_1-3 E_1^2 E_2 \eta_2-E_2^3 \eta_2)\Big]
\nonumber\\&&\quad\times\,
\Big\{2 E_1 E_2 \big[E^4+(E_1^2-E_2^2)^2-2 E^2 (E_1^2+E_2^2)\big]^2\Big\}^{-1}
\,.
\nonumber
\ea
After completing the azimuthal integration in Eq.
(\ref{normalization-condition-temp-3}) we have
\ba
&&-\,{1\over 2\pi^4}
\int {d|{\bf p}|} |{\bf p}|^2 \,\tildechiP{}^{(I)}(|{\bf p}|)^2 \, R
\nonumber\\&&
-\, {\eta_1\eta_2 E \over 8\pi^6}
\int\!\!\!\int d|{\bf p}|d|{\bf p}'|\,
|{\bf p}| |{\bf p}'|\, \tildechiP{}^{(I)}(|{\bf p}|)
\tildechiP{}^{(I)}(|{\bf p}'|) \, T^{(I)} =1 \,,
\label{normalization-2}
\ea
where $T^{(0)} = 3T(M_\rho)+T(M_\omega) + T(M_\phi)$ and
$T^{(1)} = -T(M_\rho)+T(M_\omega) + T(M_\phi)$ with
\ba
T(M_V)&=& g_{_{KK{\rm V}}}^2
\left[{2\Lambda^2-M_{\rm V}^2\over 2\Lambda^2+(|{\bf p}|+|{\bf p}'|)^2}
- {2\Lambda^2-M_{\rm V}^2\over 2\Lambda^2+(|{\bf p}|-|{\bf p}'|)^2}
\right.
\nonumber\\&&\left.\qquad\quad
+\ln{2\Lambda^2+(|{\bf p}|-|{\bf p}'|)^2
  \over 2\Lambda^2+(|{\bf p}|+|{\bf p}'|)^2}
-\,\ln{M_{\rm V}^2+(|{\bf p}|-|{\bf p}'|)^2
  \over M_{\rm V}^2+(|{\bf p}|+|{\bf p}'|)^2}
\right]
\nonumber
\ea
for each vector meson with mass $M_{\rm V}$ and coupling $g_{_{KK{\rm V}}}$\,.

If the wave function obtained in the previous section (which will be calculated
numerically in the following section) dose not satisfy this normalization
equation but gives some constant $c^2\neq 1$ for the expression on
the left hand side of Eq. (\ref{normalization-2}), one need only make the
replacement $\tildechiP(|{\bf p}|)\to\tildechiP(|{\bf p}|)/|c|$ to ensure
the correct normalization of the BS wave functions.

If the molecular binding is dominant in the $K\ovl K$ system, then the
possible bound states of the $K\ovl K$ system are most likely related to
the two particles which are denoted by $f_0(980)$ and $a_0(980)$ in the
review of PDG \cite{pdg_2006}, since they are just below the threshold of
the free $K\ovl K$ system and, up to now, can not be assigned with the
common $q\bar q$ structure. One possibility is that they are (mainly)
molecular states of other conventional particles, e.g. $K$ and $\ovl K$.
However, to identify the possible molecules of $K\ovl K$ with these scalar
particles, we should also identify other properties (other than the binding
energy) of the molecules with those of the scalar particles measured by
experiments. Among these properties, an important one is the decay width
of the bound state.

Now, we will proceed to study the decay widths of the $K\ovl K$ bound states
and compare them with those of $f_0(980)$ and $a_0(980)$ and see whether the
assignment of the molecular states with them is suitable.  Since
the dominant decay channels of $f_0(980)$ and $a_0(980)$ are $\pi\pi$
and $\eta\pi$\,, respectively, we will study the decay widths of the above
bound states into $\pi\pi$ and $\eta\pi$. The relevant interaction vertices
are (see e.g. Ref. \cite{zcsz-prd-2006})
\ba
{\cal L}_{\pi K K^*}
&=& ig_{_{\pi KK^*}} \Big[\partial_\mu K^\dag (\vec\tau\cdot \vec\pi)  K^{*\mu}
- K^\dag \,(\vec\tau\cdot \partial_\mu\vec\pi) K^{*\mu}\Big]+\hbox{c.c.} \,,
\\[5pt]
{\cal L}_{\eta K K^*}
&=& ig_{_{\eta KK^*}} \Big[\partial_\mu  K^\dag K^{*\mu}\eta
- K^\dag K^{*\mu}\partial_\mu\eta  \Big] + \hbox{c.c.} \,,
\ea
where c.c. denotes the complex conjugate of the previous terms. The coupling
constants are related to $g_{\rho\pi\pi}$ in the following way:
\ba
g_{_{\pi KK^*}} = g_{\rho\pi\pi}/2\,,\qquad
g_{_{\eta KK^*}} = - \sqrt{3} g_{\rho\pi\pi}/2 \,.
\ea
The differential decay width of the bound state can be written as
\cite{pdg_2006}
\ba
d\Gamma = {1\over 32\pi^2} |{\cal M}|^2 {|{\bf q}|\over E^2} d\Omega \,,
\label{partial-decay-width}
\ea
where $|{\bf q}|$ is the norm of the three-momentum of the
particles in the final state in the rest frame of the bound state.
${\cal M}$ is the Lorentz invariant decay amplitude of the process. The
lowest order decay amplitude can be written as (the decay to $\pi\eta$
will be considered later on)
\ba
&&\langle \pi^a(q_1)\pi^b(q_2) |\, {i^2\over 2!}
\int d^4x d^4 y\,{\rm T}\,\{ {\cal L}_{\pi K K^*}(x)
{\cal L}_{\pi K K^*}(y)\,\}  |P\rangle
\nonumber\\&&\quad
={g_{_{\pi KK^*}}^2 \over \sqrt{2E_{\pi^a}2E_{\pi^b}}}
\int {d^4 p\over (2\pi)^4}
\, \Delta_{\mu\nu}(k,M_{K^*})i^2(k+2q_2)^\mu(k-2q_1)^\nu  F(|{\bf k}|)^2
\nonumber\\ && \qquad\quad \times\,
\Big\{ (\tau^a\tau^b)_{ij}\Big|_{k=q+p+(\eta_1-\eta_2)P}
+ (\tau^b\tau^a)_{ij}\Big|_{k=q-p} \Big\}  C_{(I)}^{ji} \chiP{}^{(I)}(p)
\nonumber\\ && \qquad\quad \times\,
(2\pi)^4\delta^4(P-q_1-q_2)\,,
\label{eq: decay-amplitude-temp-1}
\ea
where $q_i$ ($i=1,2$) is the momentum of the $i$-th particle in the
final state and $E_{\pi^a} = \sqrt{{\bf q}_1^2 + m_{\pi^a}^2}$\,,
$E_{\pi^b} = \sqrt{{\bf q}_2^2 + m_{\pi^b}^2}$\,.
The coefficients $C_{(I)}^{ij}$ in Eq. (\ref{eq: decay-amplitude-temp-1})
are the isospin factors which have been given in Eqs.
(\ref{isoscalar-coefficient-0}) and (\ref{isovector-coefficient-1}),
$q\equiv \eta_2 q_1 - \eta_1 q_2$ which is not the relative momentum
of particles in the final state (note that $\eta_1$ and $\eta_2$ are defined as
$\eta_i = m_i/(m_1+m_2)$\,, and $m_1$ and $m_2$ are the masses of the
component particles of the bound states but not of the final states).
In deriving the above equation the following propagator for the vector kaons
has been used:
\ba
\langle 0 | {\rm T}\,\{K_i^{*\mu}(x)  K_j^{*\nu}(y){}^\dag\} |0\rangle
= \int {d^4k\over (2\pi)^4} \, e^{-ik(x-y)}
\Delta_{\mu\nu}(k,M_{K^*}) \delta_{ij} \,,
\ea
where $i$ and $j$ are isospin indices.

The Lorentz-invariant decay amplitude of the $K\ovl K$ bound state to
$\pi\pi$ is then
\ba
{\cal M}_{(I)}^{(\pi\pi)}&\makebox[0pt]{=}&
i c_{(I)}^{(\pi\pi)} g_{_{\pi KK^*}}^2 \sqrt{2E} \int {d^4 p\over (2\pi)^4}
\, \Big\{ \pm F(|{\bf k}|)^2\Delta_{\mu\nu}
(k,M_{K^*})i^2(k+2q_2)^\mu(k-2q_1)^\nu \Big|_{k=q-p}
\nonumber\\ &\makebox[0pt]{}& \qquad \qquad
 + (k\to q+p+(\eta_1-\eta_2)P) \Big\} \chiP{}^{(I)}(p)  \,,
\label{decay-amplitude-pi-pi}
\ea
where ``$+$'' and ``$-$'' in ``$\pm$'' are for the iso-scalar and iso-vector
channels, respectively. For the iso-scalar channel, from Eq.
(\ref{isoscalar-coefficient-0}) we have
$(\tau^a\tau^b)_{ij}C_{(0)}^{ji}={1\over\sqrt 2}{\rm Tr}(\tau^a\tau^b)
=\sqrt{2}\delta^{ab}$\,.
Since the iso-scalar $\pi\pi$ final states reads
       \footnote{
     For the states in isospin multiplets, we use the same conventions
     as those in Ref. \cite{neville-pr-67}\,. That is to say, we
     have $|\pi,\pm 1\rangle = \mp |\pi^\pm\rangle
     = \mp {1\over \sqrt 2} |\pi^1\pm i\pi^2\rangle$ and
     $|\pi,0\rangle = |\pi^3\rangle$\,.
       }
\ba
|\pi\pi\rangle_{(0,0)} = -{1\over\sqrt{3}}\left|
\pi^+\pi^- + \pi^-\pi^+ + \pi^0\pi^0 \right\rangle
= - {1\over\sqrt{3}}\left|
\pi^1\pi^1 + \pi^2\pi^2 + \pi^3\pi^3 \right\rangle \,,
\ea
we have $c_{(0)}^{(\pi\pi)} = -\sqrt{6}$\,. For the iso-vector channel,
using Eq. (\ref{isovector-coefficient-1}) we have $c_{(1)}^{(\pi\pi)}=2$
since the iso-vector $\pi\pi$ final state is
\ba
|\pi\pi\rangle_{(1,\pm 1)}&=&
- {1\over\sqrt{2}}\left|\pi^\pm\pi^0 - \pi^0\pi^\pm \right\rangle
= {1\over2}\left|\pi^0\pi^1 - \pi^1\pi^0
\pm i\pi^0\pi^2 \mp i\pi^2\pi^0\right\rangle \,,
\\
|\pi\pi\rangle_{(1,0)}&=& -  {1\over\sqrt{2}}\left|\pi^+\pi^- -
\pi^-\pi^+ \right\rangle = {i\over\sqrt{2}}\left| \pi^1\pi^2 -
\pi^2\pi^1 \right\rangle \,. \ea Note that this isospin
coefficient is independent of the component $I_3$ (notice that our
convention for the kaon state is different from that in Refs.
\cite{ldhs-npa-90,kls-prd-2004}). This iso-vector final state is
anti-symmetric, so there is in fact no S-wave $\pi\pi$ final state
with $I=1$\,.

Now, let us turn to $\pi\eta$ final state. The lowest order matrix element
for the decay of the $K\ovl K$ system into $\pi\eta$ is
\ba
&&\langle \pi^a(q_1)\eta(q_2)|2{i^2\over 2!}
\int d^4x d^4y\,{\rm T}\,\{{\cal L}_{\pi K K^*}(x)
{\cal L}_{\eta K K^*}(y)\}|P\rangle
\nonumber\\ &&
= {g_{_{\pi KK^*}}g_{_{\eta KK^*}}\over \sqrt{2E_{\pi^a}2E_{\eta}}}
(\tau^a)_{ij}C_{(I)}^{ji}
\int {d^4p\over (2\pi)^4} \, \Big\{ F(|{\bf k}|)^2
\Delta_{\mu\nu}(k,M_{K^*})i^2(k+2q_2)^\mu(k-2q_1)^\nu\Big|_{k=q-p}
\nonumber\\ && \qquad \qquad
+ (k\to q+p+(\eta_1-\eta_2)P)\Big\}\chiP{}^{(I)}(p)
\,(2\pi)^4\delta^4(P-q_1-q_2) \,.
\nonumber
\ea
Then, the Lorentz-invariant decay amplitude is (only the iso-vector channel
contributes)
\ba
{\cal M}_{(1)}^{(\pi\eta)}&\makebox[0pt]{=}&
i c_{(1)}^{(\pi\eta)} g_{_{\pi KK^*}}g_{_{\eta KK^*}} \sqrt{2E}
\nonumber\\ &\makebox[0pt]{}& \quad \times\,
\int {d^4 p\over (2\pi)^4}
\, \Big\{F(|{\bf k}|)^2\Delta_{\mu\nu}
(k,M_{K^*})i^2(k+2q_2)^\mu(k-2q_1)^\nu \Big|_{k=q-p}
\nonumber\\ &\makebox[0pt]{}& \qquad\qquad
 +\, (k\to q+p+(\eta_1-\eta_2)P) \Big\} \chiP{}^{(1)}(p)  \,.
\label{decay-amplitude-eta-pi}
\ea
The components of $\pi\eta$ are
\ba
|\pi\eta\rangle_{(1,\pm 1)}=\mp |\pi^\pm\eta\rangle
=\mp {1\over\sqrt{2}}\left|(\pi^1\pm i \pi^2)\eta \right\rangle\,,\qquad
|\pi\eta\rangle_{(1,0)}=|\pi^0\eta\rangle = |\pi^3\eta\rangle  \,.
\ea
Then, from Eq. (\ref{isovector-coefficient-1}), we have
$ c_{(1)}^{(\pi\eta)} = \sqrt{2} $ which is again independent of $I_3$\,.

In the calculation we stay in the rest frame of the bound state
and hence $P=(E,{\bf 0})$\,. In this frame the momenta of the two
particles in the final state can be taken as:
$q_1=(E_1', {\bf q}),\,  q_2 = (E_2', -{\bf q})$\,. Therefore,
$q = \eta_2 q_1-\eta_1 q_2=(\eta_2 E_1'-\eta_1 E_2',\, {\bf q})$\,.
When the final state is $\pi\pi$\,, $E_1'=E_{\pi^a}$ and $E_2'=E_{\pi^b}$
while when the final state is $\pi\eta$\,, $E_1'=E_{\pi^a}$ and
$E_2'=E_{\eta}$\,. To calculate the amplitude, we first carry out the azimuthal
integration of the spatial part of $p$, the result having the following
structure:
\ba
&&\int {d^4 p\over (2\pi)^4} \,F(|{\bf k}|)^2
\Delta_{\mu\nu}(k,M_{K^*})i^2(k+2q_2)^\mu(k-2q_1)^\nu \chiP{}^{(I)}(p)
\nonumber\\ && \qquad
= {-i^2\over (2\pi)^3}\int_{-\infty}^\infty dp^0
\int_0^\infty d|{\bf p}||{\bf p}|^2\,
f(p^0)\, \chiP{}^{(I)}(\pm p^0,|{\bf p}|) \,,
\label{eq: decay-amplitude-temp-2}
\ea
where
\ba
f(p^0)&=&
{(M_{K^*}^2-2\Lambda^2)^2\over
( - 2p^0q^0+|{\bf p}|^2 + |{\bf q}|^2+s_4 + 2\Lambda^2 +i\epsilon)^2}
\nonumber\\[5pt]&& \hspace{-2cm}\times
\left\{ 2 {(- 2p^0q^0+|{\bf p}|^2 + |{\bf q}|^2+s_4 + 2\Lambda^2)
(- 2p^0q^0+|{\bf p}|^2 + |{\bf q}|^2- s_1 + s_2s_3/M_{K^*}^2+ 2\Lambda^2)
\over \big[(|{\bf p}|-|{\bf q}|)^2+2\Lambda^2\big]
\big[(|{\bf p}|+|{\bf q}|)^2+2\Lambda^2\big]  }
\right.
\nonumber\\[5pt]&& \left.
+\,{s_1+s_4-s_2 s_3/M_{K^*}^2\over 2|{\bf p}||{\bf q}|}
\ln \left[{ 2p^0q^0+2|{\bf p}||{\bf q}|-s_4
 \over  2p^0q^0-2|{\bf p}||{\bf q}|-s_4 }
\cdot {(|{\bf p}|-|{\bf q}|)^2+2\Lambda^2
\over (|{\bf p}|+|{\bf q}|)^2+2\Lambda^2}
\right]\right\} \,.
\nonumber
\ea
Now we will give some explanations about Eq.
(\ref{eq: decay-amplitude-temp-2}). The results for $k=q-p$ and
$k=q+p+(\eta_1-\eta_2)P$ have the same structures.
We have changed the sign of the imaginary part of the pole for
$k=q+p+(\eta_1-\eta_2)P$ by taking the variable transformation $p^0\to -p^0$\,.
When $k=q-p$ we will take $\chiP{}^{(I)}(+ p^0,|{\bf p}|)$ in Eq.
(\ref{eq: decay-amplitude-temp-2}) and $s_i$ ($i=1,\dots,4$) are defined by
\ba
s_1&=& p^2 + q^2 + 4(\eta_1^2-\eta_1) P^2 + 2(2\eta_1-1)(p\cdot P + q\cdot P)
\,, \nonumber\\
s_2&=& p^2 - q^2 + 2\eta_1(p\cdot P - q\cdot P) \,,
\nonumber\\
s_3&=& p^2 - q^2 + 2(\eta_1-1)(p\cdot P -q\cdot P) \,,
\nonumber\\
s_4&=& p^2 + q^2 - M_{K^*}^2 \,, \nonumber \ea while when
$k=q+p+(\eta_1-\eta_2)P$ we will take $\chiP{}^{(I)}(-p^0,|{\bf
p}|)$ in Eq. (\ref{eq: decay-amplitude-temp-2}) and $s_i$
($i=1,\dots,4$) are defined by \ba s_1&=& p^2 + q^2 - P^2 \,,
\nonumber\\
s_2&=& p^2 - q^2 - (2\eta_1-1)P^2 - 2(\eta_1-1)p\cdot P - 2\eta_1q\cdot P \,,
\nonumber\\
s_3&=&  p^2 - q^2 + (2\eta_1-1) P^2 - 2\eta_1p\cdot P - 2(\eta_1-1)q\cdot P \,,
\nonumber\\
s_4&=& p^2 + q^2 - M_{K^*}^2 + (2\eta_1-1)^2 P^2 -
2(2\eta_1-1)p\cdot P + 2(2\eta_1-1)q\cdot P \,. \nonumber \ea Now,
we can substitute Eq. (\ref{4-dim-BS}) into Eq. (\ref{eq:
decay-amplitude-temp-2}) and complete the $p^0$-integration by
choosing proper contours. From the expression of $f(p^0)$ above we
can see that the all the poles come from the BS wave function.
This is because the denominator in $f(p^0)$ (neglecting the
isospin violation, then $\eta_1=\eta_2$) \ba - 2p^0q^0+|{\bf p}|^2
+ |{\bf q}|^2+s_4 + 2\Lambda^2 = (p^0-q^0)^2+2\Lambda^2-M_{K^*}^2
\nonumber \ea is positive when $\Lambda>M_{K^*}/\sqrt{2}$, which
is satisfied in our case (see the discussion in the next section).
The remaining contour integration over $p^0$ is straightforward
and the result reads \ba &&\int {d^4 p\over (2\pi)^4}\, \Big\{
F(|{\bf k}|)^2\,
\Delta_{\mu\nu}(k,M_{K^*})i^2(k+2q_2)^\mu(k-2q_1)^\nu
\Big(\Big|_{k=q-p} \pm \Big|_{k=q+p}\Big) \Big\}\chiP{}^{(I)}(p)
\nonumber\\[5pt]
&=& {i^4\over (2\pi)^3}\int_0^\infty d|{\bf p}||{\bf p}|^2
\left[ \xi_1\, f(p^0)\big|_{p^0=-\eta_1 E - E_1}
+\xi_2\, f(p^0)\big|_{p^0=\eta_2 E - E_2}
\right.\nonumber\\ &&\hspace{2cm}\left.
\pm \xi_3\, f(p^0)\big|_{p^0=\eta_1 E - E_1}
\pm \xi_4\, f(p^0)\big|_{p^0=-\eta_2 E - E_2}
\right]\tildechiP{}^{(I)}(|{\bf p}|)  \,,
\label{common-structure}
\ea
where $\xi_1 = E_2(E-E_1-E_2)/[(E_1+E_2)(E+E_1-E_2)]$,
$\xi_2 = E_1(E+E_1+E_2)/[(E_1+E_2)(E+E_1-E_2)] $\,,
$\xi_3=E_2(E+E_1+E_2)/[(E_1+E_2)(E-E_1+E_2)]$\,, and
$\xi_4 = E_1(E-E_1-E_2)/[(E_1+E_2)(E-E_1+E_2)]$\,.
If $\eta_1=\eta_2$ we have $\xi_1=\xi_4$ and $\xi_2=\xi_3$, then for the
iso-vector $\pi\pi$ final state, the decay width is zero.

Once we have obtained the BS wave function of the ground
state $\chiP(p)$ (the numerical calculation will be carried out in the next
section), we will take the wave function as input to calculate the decay
amplitudes in Eqs. (\ref{decay-amplitude-pi-pi}) and
(\ref{decay-amplitude-eta-pi}).

\section{Numerical analysis and results}
\label{sect-numerical}

The cutoff $\Lambda$ in our model is {\it not} a free parameter in
principle. It contains the information about the non-point
interaction due to the structures of hadrons. In Ref.
\cite{ldhs-npa-90}, the cutoff for the interaction of $K\ovl
K\rho$ is taken to be rather large (about 3.18 GeV in our
notation). On the other hand, in the study of baryons in the
quark-diquark picture, the cutoff in the form factors associated
with the diquark-gluon-diquark interaction is taken to be about
1.27 GeV \cite{diquark-form-factor}. In this work, we shall treat
the cutoff $\Lambda$ in the form factors as a parameter varying in
a much wider range $(0.8, 4.8)$ GeV, in which we will try to
search for possible solutions of the $K\ovl K$ bound states.

Let us first solve the BS bound state equation
(\ref{scalar-BS-eq-full-final}) numerically. We discretize the integral
equation (\ref{scalar-BS-eq-full-final}) into a matrix eigenvalue equation
by the Gaussian quadrature method. For each pair of trial values of the
cutoff $\Lambda$ and the binding energy $E_{\rm b}$ of the $K\ovl K$ system
(which is defined as $E_{\rm b} = E-m_1-m_2$), we will
obtain all the eigenvalues of this eigenvalue equation. The eigenvalue closest
to 1.0 for a pair of $\Lambda$ and $E_{\rm b}$ will be selected out and called
``the-trial-eigenvalue''. Fixing a value of the cutoff $\Lambda$ and varying
the binding energy $E_{\rm b}$ (from 0 to $-100$ MeV) we will obtain a series
of ``the-trial-eigenvalue''s. For some ({\it not} all) values of the cutoff,
we will find that the corresponding series cross over 1.0
       \footnote{
     That is to say, e.g., from 0.99 to 1.01\,.
       }
in the range of $E_{\rm b}\in (0,-100)$ MeV. The task is then to find
out all these cutoff values (which are, in fact, some continuous regions).

In searching for the possible solutions in the iso-scalar channel of the
$K\ovl K$ system and its contribution to $f_0(980)$ ($I^G(J^{PC})=0^+(0^{++})$),
we find several regions of the cutoff. The results are listed in Table
\ref{table: isoscalar}.
\begin{table}
\caption{\label{table: isoscalar} For the iso-scalar $K\ovl K$
system, there are five regions of the cutoff $\Lambda$\,. In each
region, for any value of the cutoff, the series of
``the-trial-eigenvalue''s cross over the exact eigenvalue $1.0$
(at certain binding energy $E_{\rm b}\in [-1,-99]$ MeV). } \center
\begin{tabular}{c|ccccc}
\hline
$E_{\rm b}$ ( MeV) &  & &$\Lambda$ (GeV) & &  \\
\hline
$-1$ & 1.1360 & 2.0793 & 2.7352 & 3.5453 & 4.7633 \\
\hline
$-99$ & 1.2162 & 2.0979 & 2.7444 & 3.5524 & 4.7697\\
\hline
\end{tabular}
\end{table}
If we want to identify the iso-scalar bound state of the $K\ovl K$ system
with the observed $f_0(980)$\,, we should let the binding energy approximately
be $-20$ MeV. For the above five regions, this is
equivalent to set the cutoff to be
\ba
\Lambda \approx 1.1700 \,,\quad 2.0862 \,,\quad 2.7385 \,,\quad 3.5479 \,,\quad
 4.7656 \hbox{\quad GeV} \,,
\ea
respectively.
From Eqs. (\ref{decay-amplitude-pi-pi}) and (\ref{partial-decay-width}),
taking the corresponding BS wave function as input, the decay width
of the iso-scalar $K\ovl K$ to $\pi\pi$ can be obtained. The results are listed
in Table \ref{table: isoscalar-decay-width}.
From the PDG's review \cite{pdg_2006} the full width of
$f_0(980)$ is $\Gamma =40\hbox{ to }100\hbox{ MeV}$ ($\pi\pi$ dominant).
Therefore, the results in Table \ref{table: isoscalar-decay-width} show
that $f_0(980)$ can not be completely the $K\ovl K$ iso-scalar bound state.
\begin{table}
\caption{\label{table: isoscalar-decay-width} The decay widths
($\Gamma_{\pi\pi}^{(I=0)}$) corresponding to the five cutoff-regions when the
binding energy $E_{\rm b} \approx -20$ MeV.
}
\center
\begin{tabular}{c|ccccc}
\hline
$\Lambda$ (GeV) & 1.1700 & 2.0862 & 2.7385 & 3.5479 & 4.7656 \\
\hline
$\Gamma$ (MeV) & 0.671 & 4.758 & 7.516 & 11.102 & 11.949 \\
\hline
\end{tabular}
\end{table}

Now, let us turn to the iso-vector $K\ovl K$ bound state. We find the following
two regions of the cutoff in this case:
\ba
\Lambda\in (2.1160,2.2213)\quad \hbox{and}\quad (4.4998,4.5147)\quad\hbox{GeV}
\,.
\label{isovector-cutoff-region}
\ea
The corresponding decay widths of the $K\ovl K$ system into $\pi\eta$ with
the binding energy $E_{\rm b} = -20$ MeV are given by
\ba
\Gamma_{\pi\eta}^{(I=1)} &=& 1.329 \hbox{\quad MeV}\,,\quad
\Lambda = 2.1590 \hbox{\quad GeV}  \,,
\\
\Gamma_{\pi\eta}^{(I=1)} &=& 0.031 \hbox{\quad MeV}\,,\quad
\Lambda = 4.5056 \hbox{\quad GeV} \,.
\ea
%
%
%
%
The full width of $a_0(980)$ is $\Gamma =50\hbox{ to }100\hbox{
MeV}$ ($\pi\eta$ dominant)\cite{pdg_2006}. Although the $K\ovl K$
iso-vector bound state does contribute to $a_0(980)$
($I^G(J^{PC})=1^-(0^{++})$), just as in the case of $f_0(980)$,
$a_0(980)$ can not be completely the $K\ovl K$ iso-vector bound
state.

\section{Conclusions and discussions}

In this paper we derive the BS equation for the $K\ovl K$ system,
study the possible bound states of this system, and calculate
their decay widths in the BS formalism. In our model, we have used
the ladder approximation. This approximation has been questioned
and is found not to be a good one in some models where higher
order graphs give even more important contribution than the ladder
graph
\cite{nieuwenhuis-tjon-prl-77-814,theussl-desplanques-fbs-30-5,barrobergflodt-etal-fbs-2006,emamirazavi-etal-jpg-2006}.
However, in our case, we have shown explicitly that crossed-ladder
graphs are suppressed greatly comparing with the ladder graphs due
to the large masses of the exchanged particles. This makes the
ladder approximation be legitimate in our model. In addition,
based on the fact that the $K\ovl K$ system is weakly bound, we
have used the instantaneous approximation in the BS equation, in
which the energy exchange between the constituent particles is
neglected. Since the constituent particles and the exchanged
particles in the $K\ovl K$ system are not point-like, we introduce
a form factor including a cutoff $\Lambda$ which reflects the
effects of the structure of these particles. Since $\Lambda$ is
controlled by non-perturbative QCD and can not be determined at
present, we let it vary in a reasonable range within which we try
to find possible bound states of the $K\ovl K$ system.

From the calculating results we find that there exist bound states of the
$K\ovl K$ system. Unfortunately, we can not determine the binding energy
uniquely. The binding energy depends on the value of the cutoff $\Lambda$\,.
For the iso-scalar $K\ovl K$ system, we find five cutoff regions in which the
solutions (with the binding energy $E_{\rm b}\in (0,-100)$ MeV) to the ground
state of the BS equation can be found (in unit of GeV):
\ba
\Lambda \sim (1.136,1.216)\,,~(2.079,2.098)\,,~(2.735,2.744)\,,
~(3.545,3.552)\,,~ (4.763,4.770) \,.
\nonumber
\ea
From these results, we can see that, except for the first interval, these
regions are very narrow. For the iso-vector case, we find two regions
(in unit of GeV),
\ba
\Lambda \sim (2.1160,2.2213)\,, \quad (4.4998,4.5147) \,.
\nonumber
\ea

How to fix the cutoff (then the binding energy can be predicted), which is
equivalent to how to determine the finite size effects of hadrons in the
calculation, is beyond the scope of this paper.
If we treat the binding energy as an input ($E_{\rm b} = -20 $ MeV),
we find that the corresponding BS wave function gives too small decay widths,
i.e.
\ba
\Gamma_{\pi\pi}^{(I=0)} &=& 0.671 \,,~ 4.758 \,,~ 7.516 \,,~ 11.102 \,,~ 11.949
\hbox{\quad MeV} \,,
\nonumber
\ea
corresponding to the five cutoff regions, respectively. For the iso-vector
case we have
\ba
\Gamma_{\pi\eta}^{(I=1)} &=& 1.329\,,~ 0.0305 \hbox{\quad MeV} \,,
\nonumber
\ea
corresponding to the two cutoff regions, respectively.

The authors in Ref. \cite{kls-prd-2004} concluded that the model with the
one-meson-exchange potential from chiral dynamics, which is also used in this
work, is sufficient to bind the $K\ovl K$ system into a molecule which has the
same mass and decay width as those of the iso-scalar $f_0(980)$. From our
calculation, however, we find that even these ($K\ovl K$) bound states could
contribute to the observed scalar particles, the portion should be small
      \footnote{
    The numerical results show that when the binding is stronger (i.e., for
    larger $|E_{\rm b}|$) the decay widths will become larger.
      }.
We prefer to draw the conclusion that there may be some more important
structures besides the $K\ovl K$ molecule in the observed overpopulated
scalar particles (e.g. $f_0(980)$ and $a_0(980)$). Obviously, to resolve this
problem further investigations are required.


\bigskip
\bigskip

\noindent
{\bf Acknowledgments.}
One of us (XHW) is grateful to Dr. Wei Zhang for the help on Fortran
programing. This work was supported in part by National Natural Science
Foundation of China (Project Number 10675022), the Key Project of Chinese
Ministry of Education (Project Number 106024) and the Special Grants from
Beijing Normal University.


\begin{thebibliography}{99}

  \bibitem{gn-rmp-99}
    S. Godfrey and J. Napolitano, 
    Rev. Mod. Phys. {\bf 71}, 1411 (1999).
  \bibitem{jaffe-prd-77}
    R.L. Jaffe,
    Phys. Rev. {\bf D15}, 267 (1977).
  \bibitem{ai-npb-89}
    N.N. Achasov and V.N. Ivanchenko,
    Nucl. Phys. {\bf B315}, 465 (1989).
  \bibitem{as-prd-98}
    N.N. Achasov and G.N. Shestakov,
    Phys. Rev. {\bf D58}, 054011 (1998).
  \bibitem{a-ph-0309118}
    N.N. Achasov,
    Phys. Atom. Nucl. {\bf 67} (2004) 1529, Yad. Fiz. {\bf 67} (2004) 1552.
  \bibitem{wi-prd-83-90}
    J. Weinstein and N. Isgur,
    Phys. Rev. {\bf D27}, 588 (1983) ;\\
    J. Weinstein and N. Isgur, 
    Phys. Rev. {\bf D41}, 2236 (1990).
  \bibitem{ldhs-npa-90}
    D. Lohse, J.W. Durso, K. Holinde, and J. Speth,
    Nucl. Phys. {\bf A516}, 513 (1990).
  \bibitem{jphs-prd-95}
    G. Janssen, B.C. Pearce, K. Holinde, and J. Speth,
    Phys. Rev. {\bf D52}, 2690 (1995).
  \bibitem{oller-npa-2003}
    J.A. Oller, Nucl. Phys. {\bf A714}, 161 (2003).
  \bibitem{kls-prd-2004}
    S. Krewald, R.H. Lemmer, and F.P. Sassen,
    Phys. Rev. {\bf D69}, 016003 (2004).
  \bibitem{zcsz-prd-2006}
    Y.-J. Zhang, H.-C. Chiang, P.-N. Shen, and B.-S. Zou,
    Phys. Rev. {\bf D74}, 014013 (2006).
  \bibitem{a-ex-98-00}
    N.N. Achasov et al., Phys. Lett. {\bf B440}, 442 (1998),
    ibid. {\bf B438}, 441 (1998),
    ibid. {\bf B479}, 53 (2000),
    ibid. {\bf B485}, 349 (2000).
  \bibitem{kloe-2002}
    KLOE Collaboration (A. Aloisio et al.), Phys. Lett. {\bf B536}, 209 (2002),
    ibid. {\bf B537}, 21 (2002).
  \bibitem{belle-cdfII-d0-babar}
    Belle Collaboration (S.K. Choi, et al.), Phys. Rev. Lett. {\bf 91}, 262001 (2003); Belle Collaboration (K. Abe, et al.), hep-ex/0308029v1;\\
    CDF II Collaboration (D. Acosta et al.), Phys. Rev. Lett. {\bf 93}, 072001 (2004);\\
    D0 Collaboration (V.M. Abazov et al.), Phys. Rev. Lett. {\bf 93}, 162002 (2004);\\
    BABAR Collaboration (B. Aubert et al.), Phys. Rev. {\bf D71}, 071103 (2005).


  \bibitem{heavy-molecules}
    T. Barnes, F.E. Close, and H.J. Lipkin, Phys. Rev. {\bf D68}, 054006 (2003); \\
    X. Liu, X.-Q. Zeng, and X.-Q. Li, Phys. Rev. {\bf D72}, 054023 (2005)

  \bibitem{gross-prc-26-2203}
    F. Gross, Phys. Rev. {\bf C26}, 2203 (1982).

  \bibitem{itzykson-zuber-books}
    C.C. Itzykson and J.-B. Zuber,  ``{\it Quantum Field Theory}'', Vol. 2, Chapter 10,
    (McGraw-Hill, New York, 1985).

  \bibitem{nieuwenhuis-tjon-prl-77-814}
    T. Nieuwenhuis and J.A. Tjon, Phys. Rev. Lett. {\bf 77}, 814 (1996).

  \bibitem{theussl-desplanques-fbs-30-5}
    L. Theu{\ss}l and B. Desplanques, Few-Body Systems {\bf 30}, 5 (2001).


  \bibitem{barrobergflodt-etal-fbs-2006}
    K. Barro-Bergfl\"odt, R. Rosenfelder, and M. Stingl, Few-Body Systems
    {\bf 39}, 193 (2006).

  \bibitem{emamirazavi-etal-jpg-2006}
    M. Emami-Razavi1 and J.W. Darewych, J. Phys. G: Nucl. Part. Phys.
    {\bf 32}, 1171 (2006).

  \bibitem{lurie-book}
    David Lurie, ``{\it Particles and Fields}'', Chapt. 9,
    (Interscience Publishers, 1968).
  \bibitem{neville-pr-67}
    D.E. Neville, Phys. Rev. {\bf 160}, 1375 (1967).
  \bibitem{KSRF-66}
    K. Kawarabayashi and M. Susuki,
    Phys. Rev. Lett. {\bf 16}, 255 (1966) ;\\
    X. Riazuddin and X. Fayyazuddin,
    Phys. Rev. {\bf 147}, 1071 (1966).

  \bibitem{diquark-form-factor}
    M. Anselmino, P. Kroll, and B. Pire, Z. Phys. {\bf C36}, 89 (1987);\\
    X.-H. Guo and T. Muta, Phys. Rev. {\bf D54}, 4629 (1996);\\
    X.-H. Guo, A.W. Thomas, and A.G. Williams, Phys. Rev. {\bf D59}, 116007 (1999).

  \bibitem{pdg_2006}
    Particle Data Group (W.-M. Yao et al.),
    J. Phys. {\bf G33}, 1 (2006).

\end{thebibliography}
\end{document}